**Influence of the gate dielectric on the mobility of rubrene single-crystal field-effect transistors**

A. F. Stassen, R.W.I. de Boer, N.N. Iosad, and A.F. Morpurgo.

*Kavli Institute of Nanoscience, Delft University of Technology, Lorentzweg 1, 2628 CJ DELFT, The Netherlands*

We have performed a comparative study of rubrene single-crystal field-effect transistors fabricated using different materials as gate insulator. For all materials, highly reproducible device characteristics are obtained. The achieved reproducibility permits to observe that the mobility of the charge carriers systematically decreases with increasing the dielectric constant of the gate insulator, the decrease being proportional to $\varepsilon^{-1}$. This finding demonstrates that the mobility of carriers in organic single-crystal field-effect transistors is an intrinsic property of the crystal/dielectric interface and that it does not only depend on the specific molecule used.

PACS numbers: 71.20.Rv, 70.80.Le, 73.40.Qv.

Recent research effort has led to the successful fabrication of field-effect transistors at the surface of single crystals of organic molecules. Work performed by different groups has resulted in single-crystal devices of very high quality, exhibiting an unprecedented level of reproducibility. For molecules such as tetracene, pentacene and rubrene, essentially identical results (e.g., comparable values for charge carrier mobility) have been obtained in different laboratories and using different device fabrication techniques.[1-6]

The quality of organic single-crystal FETs opens new opportunities for investigations of both fundamental and applied character. In particular, the use of single-crystalline devices permits to study the intrinsic –not limited by disorder- transport properties of organic semiconductors as a function of carrier density, as recently demonstrated by the observation of an anisotropic mobility in rubrene FETs exhibiting a "metallic-like" temperature dependence.[7, 8] In addition, the reproducibility of single-crystal FETs permits to investigate in detail how different aspects of the devices influence transistor operation, which is necessary to individuate the ultimate performance limits of organic transistors.

In this paper, we report a comparative experimental study of the electrical characteristics of rubrene single-crystal FETs fabricated using $Ta_2O_5$, $Al_2O_3$, $SiO_2$, and Parylene C as gate insulator. For the different dielectrics, field-effect transistors exhibiting stable and hysteresis-free electrical behavior can be reproducibly realized. In all cases, the hole mobility extracted from room-temperature measurement of the transistor characteristics is remarkably gate-voltage independent. From these measurements, we find that the mobility decreases from 10 $cm^2$/Vs (Parylene C, $\varepsilon=3.15$) to 1.5 $cm^2$/Vs ($Ta_2O_5$, $\varepsilon=25$) with increasing the relative dielectric constant. By comparing our data to those recently reported for transistors fabricated using Parylene N ($\mu = 15$ $cm^2$/Vs; $\varepsilon= 2.65$)[2, 9] and PDMS air-gap stamps ($\mu=20$ $cm^2$/Vs; $\varepsilon=1$)[7], we conclude that a decrease in mobility with increasing the dielectric constant of the gate dielectric occurs systematically in rubrene single-crystal FETs. This result demonstrates that the mobility measured in organic transistors is not only a property of the specific organic molecule used, but that it intrinsically depends on the organic/dielectric interface.

The devices used in our investigations have been fabricated by means of two different, recently developed techniques. Transistors based on Parylene C have been built following the processing described in Ref. [1], using aqueous colloidal graphite or silver epoxy for the source, drain, and gate electrodes (with colloidal graphite resulting in better performances as compared to epoxy). For these devices, rather thick crystals (typically 100 μm or thicker) were used. All other transistors were fabricated by means of electrostatic bonding of much thinner (approximately 1 micron thick) crystals on doped silicon substrates with pre-fabricated FET circuitry, using a process identical to the one described in Ref.[3]. In all cases, the crystals were grown in a horizontal oven similar to the system used in Ref.[10] with argon as carrier gas (50 ml/min). The majority of the rubrene crystals used in our investigations were needle shaped. As discussed in [7], these needle-shaped crystals grow preferentially along the crystallographic *b*-axis, which corresponds to the direction of highest hole mobility[7]. Both the parylene gate and the oxide-gate transistors were fabricated with a large source-drain distance (always larger than 300 μm and typically ~ 1 mm) to ensure that the measured transport occurred preferentially along this direction and to minimize contact effects on measurements performed in a two-terminal configuration.

In the case of $SiO_2$-based transistors, the gate insulator was a 200 nm thick, thermally grown oxide layer. In the case of $Ta_2O_5$, a Nb layer acting as a gate was sputter-deposited onto the Si substrate, followed by a 375 nm thick layer of $Ta_2O_5$. Sputtering of $Ta_2O_5$ was performed from a metallic Ta target in the presence of oxygen in an argon plasma, with the substrate held at approximately 300 C, according to the procedure developed in Ref. [11]. Contrary to the case of $Ta_2O_5$ layers sputtered from a ceramic target[12], this procedure results in negligibly low leakage current, at least up to gate fields of 3 MV/cm. finally, $Al_2O_3$ devices were fabricated by sputtering a 25 nm layer of $Al_2O_3$ on top of a $Ta_2O_5$ layer. The roughness of the three different oxide layers was measured using an atomic force microscope and found to be less than 0.1 nm. Figure 1 shows optical images of two rubrene single-crystal FETs fabricated by means of electrostatic bonding on $Al_2O_3$ (A) and on $SiO_2$ (B).

Figure 2 shows typical source-drain voltage-current $I_d$ vs. $V_{sd}$ sweeps taken at different values of the voltage applied to the gate electrode ($V_g$) for a single-crystal FET on $SiO_2$ (the inset shows similar data for a transistor fabricated using Parylene C). All measurements discussed here were performed in a two-terminal configuration in vacuum ($10^{-6}$ mbar), at room temperature, using a HP4156A semiconductor parameter analyzer. As shown in figure 2, essentially no hysteresis is present in the measurements and the $I_d$ vs. $V_{sd}$ are linear down to small applied $V_{sd}$ voltages. For electrostatically bonded transistors we performed in a number of cases four-terminal measurements and observed that owing to large length of the FET channel the contact resistance does not significantly affect the value of mobility observed in a two terminal configuration. In the case of Parylene C, the manually deposited contacts are often of lower quality and the application of a large source-drain bias is required to avoid that the current contact limitations to the current flowing through the devices.

We use the relation (W is the channel width, L the channel length, and $C_d$ the capacitance per unit area)

$$\mu = \frac{L}{W} \times \frac{1}{C_d} \times \frac{1}{V_{sd}} \times \frac{\delta I_d}{\delta V_g} \tag{1}$$

to extract the value of the mobility from the linear part of the transistor characteristics, as a function of gate voltage. Figure 3A shows the mobility data obtained using this relation for different values of the source drain voltage ($V_{sd}$ = -5, -7.5, -10, -12.5, and –15 V respectively) for one of the electrostatically bonded transistors with $SiO_2$ gate insulator. The same value of mobility is obtained independent of $V_g$ and $V_{sd}$ in all cases, as long as the device operates in the linear regime for which Eq. 1 applies, i.e. if $V_g$ is sufficiently larger than Vsd (the apparent peak in mobility present at low $V_g$ is an artifact due to the violation of this condition). The field-effect mobility extracted from the transistor *I-V* characteristics as a function of the applied gate voltage is shown in Fig. 3B for transistors fabricated using all the different gate insulators.

Figure 3B also shows that the mobility of rubrene single-crystal FETs is different for devices fabricated with the different gate dielectrics. Although all transistors fabricated show a spread in the values of measured mobilities, probably originating from defects induced by the crystal handling during the fabrication process, we found that a large fraction of devices exhibit mobilities in a rather narrow range of values. Specifically, for transistors fabricated using Parylene C $\mu$ typically ranges between 6 and 10 cm$^2$/Vs, for SiO$_2$ between 4 and 6 cm$^2$/Vs, for Al$_2$O$_3$ devices best $\mu$ values are 2-3 cm$^2$/Vs, and for Ta$_2$O$_5$ 1-1.5 cm$^2$/Vs. From these data, it is apparent that for all investigated rubrene FETs the measured mobility systematically decreases when increasing the dielectric constant of the gate insulator. This trend is consistent with the results obtained by others on rubrene FETs fabricated using Parylene N ($\varepsilon$=2.65), for which the measure mobility ranges between 10 and 15 cm$^2$/Vs [9], and vacuum ($\varepsilon$=1) where the mobility range is 16-20 cm$^2$/Vs [7].

Figure 4 summarizes the available data and clearly demonstrates the dependence of $\mu$ on the dielectric constant $\varepsilon$. The inset of Fig. 4 additionally shows that, when plotted on a log-log scale, the mobility decrease with increasing $\varepsilon$ as $\varepsilon^{-1}$ over the entire range available (slightly more than one decade). From these data we directly conclude that the mobility of organic FETs cannot be simply considered to be an intrinsic property of the molecular material used but rather that it is *intrinsically* a property of the *organic/dielectric interface*. This conclusion is also supported by our measurements on tetracene single-crystal FETs, in which we have observed that the mobility of SiO$_2$ based devices is systematically larger than that of transistors fabricated using Ta$_2$O$_5$ as a gate insulator.

A systematic decrease in $\mu$ with increasing $\varepsilon$ has recently been reported for disordered polymeric organic FETs of lower mobility[13]. In that context, the effect has been attributed to the localized nature of the charge carriers in the material and their interaction with the induced polarization in the gate insulator. Specifically, according to the authors of Ref. [13], dipolar disorder in the dielectric, which is stronger the larger the value of the dielectric constant, induces a broadening of the density of states (DOS) at the polymer/insulator interface. This broadening results in a decrease of the DOS at the Fermi energy, which, in a disordered material, causes a lower hopping probability. This lower hopping probability leads to a suppression of the carrier mobility, in agreement with theoretical work by Bassler.[14]

In single-crystalline devices, disorder is much weaker than in polymers. Nevertheless, even in the best rubrene single-crystal FETs at room temperature holes are nearly localized by polaronic effects and cannot be described in terms of extended states, as it is the case for conventional inorganic semiconductors. Therefore, also in single-crystal FETs, a (nearly) localized charge carrier at the rubrene/dielectric interface locally polarizes the dielectric. The electrostatic potential generated by the induced polarization exerts an attractive force on the charge carrier itself that increases the tendency towards carrier self-trapping. As the attractive force is larger for larger $\varepsilon$, this qualitatively explains why the mobility is reduced with increasing $\varepsilon$. Stated differently, at the interface between the crystal and the gate insulator, the electrical polarizability of the environment experienced by the charge carriers is determined by the dielectric constant of the insulating material. In this way, the polaronic dressing of charge carriers is enhanced by the presence of a gate insulator with a large dielectric constant. As a consequence of this enhanced polaronic dressing, the mobility is reduced.

The detailed microscopic understanding of the mechanism just proposed clearly requires (and deserves) further experimental and theoretical investigations. From a fundamental perspective, this mechanism is interesting, since it permits to tune polaronic effects in a FET configuration, thus offering a new tool for their study. For instance, experimentally, it will be interesting to look in detail at how the temperature dependence of the mobility evolves from the metallic-like regime ($d\mu/dT <0$) [8] to the thermally activated regime ($du/dT >0$) [15] with increasing $\varepsilon$ (work is in progress in this direction). Finally, our findings are also relevant for applications, as they clearly demonstrate that the speed of organic transistors can be enhanced by using low-$\varepsilon$ gate insulators.

We acknowledge useful discussions with D. de Leeuw, M.E. Gershenson, S. Goennenwein, T.M. Klapwijk, and J. Veres. This work was financially supported by FOM. The work of AFM is part of the NWO Vernieuwingsimpuls 2000 program.

**References**

[1]  V. Podzorov, V. M. Pudalov and M. E. Gershenson, Appl. Phys. Lett. **82** (11), 1739 (2003).

[2]  V. Podzorov, S. E. Sysoev, E. Loginova, V. M. Pudalov and M. E. Gershenson, Appl. Phys. Lett. **83** (17), 3504 (2003).

[3]  R. W. I. de Boer, T. M. Klapwijk and A. F. Morpurgo, Appl. Phys. Lett. **83** (21), 4345 (2003).

[4]  V. Y. Butko, X. Chi, D. V. Lang and A. P. Ramirez, Appl. Phys. Lett. **83** (23), 4773 (2003).

[5]  J. Takeya, C. Goldmann, S. Haas, K. P. Pernstich, B. Ketterer and B. Batlogg, J. Appl. Phys. **94** (9), 5800 (2003).

[6]  R. W. I. de Boer, M. E. Gershenson, A. F. Morpurgo and V. Podzorov, Phys. Stat. Sol. A **201** (6), 1302 (2004).

[7]  V. C. Sundar, J. Zaumseil, V. Podzorov, E. Menard, R. L. Willett, T. Someya, M. E. Gershenson and J. A. Rogers, Science **303** (5664), 1644 (2004).

[8]  V. Podzorov, E. Menard, A. Borissov, V. Kiryukhin, J. A. Rogers and M. E. Gershenson, cond-mat/, 0403575 (2004).

[9]  M. E. Gershenson (private communication).

[10]  R. A. Laudise, C. Kloc, P. G. Simpkins and T. Siegrist, J. Cryst. Growth **187** (3-4), 449 (1998).

[11]  N. N. Iosad, G. J. Ruis, E. V. Morks, A. F. Morpurgo, N. M. van der Pers, P. F. A. Alkemade and V. G. M. Sivel, J. Appl. Phys. **95** (12), 8087 (2004).

[12]  R. M. Fleming, D. V. Lang, C. D. W. Jones, M. L. Steigerwald, D. W. Murphy, G. B. Alers, Y. H. Wong, R. B. van Dover, J. R. Kwo and A. M. Sergent, J. Appl. Phys. **88**, 850 (2000).

[13]  J. Veres, S. D. Ogier, S. W. Leeming, D. C. Cupertino and S. D. Khaffaf, Adv. Mater **13** (3), 199 (2003).

[14]  H. Bässler, Philos. Mag. **50**, 347 (1984).

[15]  R. W. I. de Boer *et al.* unpublished results.

**Figures Captions**

**Figure 1.**

Two rubrene single-crystal FETs fabricated by means of electrostatic bonding on $Al_2O_3$ (A) and $SiO_2$ (B). In (A) the electrodes have been defined by evaporation through a shadow mask, whereas in (B) photo-lithography and lift-off were used. In both figures, the bar is 200 μm long.

**Figure 2.**

Source-drain current vs. source-drain voltage measured at different gate voltages for a device fabricated on $SiO_2$, with a L=1.2 mm channel lenght and W=200 μm channel width. The inset shows similar data for a FET fabricated using parylene C (L= 650 μm, W=340 μm).

**Figure 3.**

(A) Mobility vs. gate voltage for a device on $SiO_2$, measured at different values of source-drain voltage ($V_{sd}$ = -5, -7.5, -10, -12.5, -15 V, respectively), obtained using Eq. 1. Note that in the linear regime ($V_g \gg V_{sd}$) μ does not depend on $V_g$ and $V_{sd}$ (the apparent peak at low $V_g$ values is an artifact originating from the use of Eq. 1 outside the linear regime). (B) $\mu(V_g)$ curves as measured for the four different gate insulators. For device based on parylene C, the suppression of contact effects often requires the use of a rather large value $V_{sd}$ (and thus $V_g$, to remain in the linear regime).

**Figure 4.**

Decrease of the mobility with increasing ε, as observed in rubrene single-crystal FETs with different gate insulators. The bars give a measure of the spread in mobility values. Inset: when plotted on a log-log-scale, the available data show a linear dependence with slope –1 (i.e. the variation in μ is proportional to $\varepsilon^{-1}$).

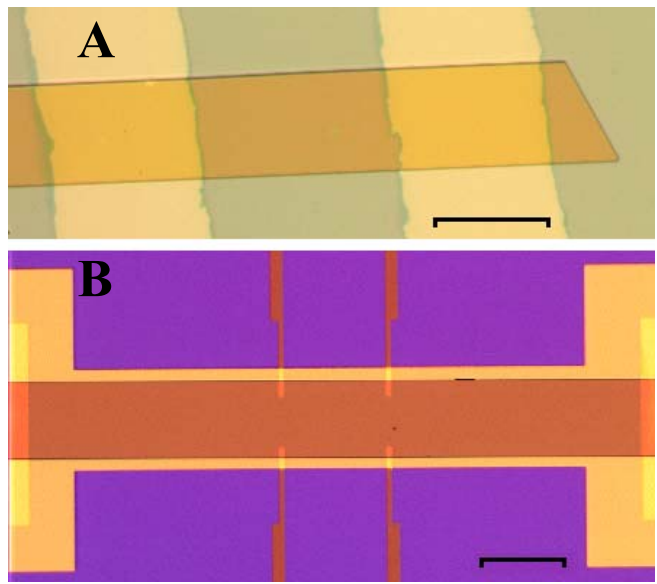

Figure 1 A. F. Stassen *et al.*

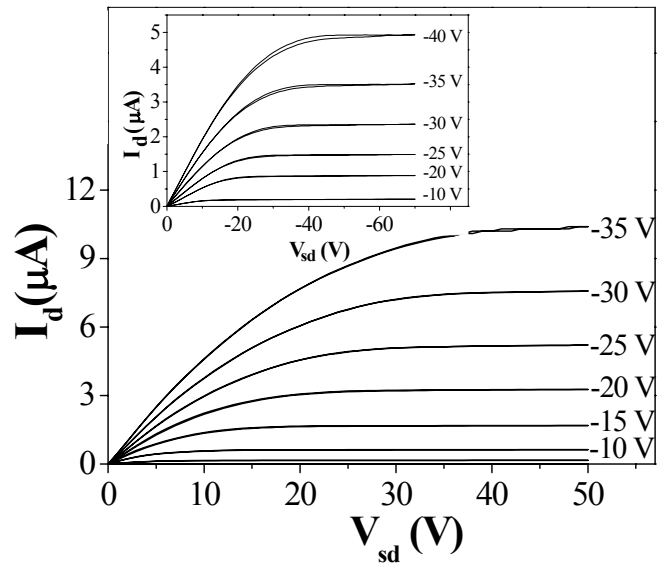

Figure 2 A. F. Stassen *et al.*

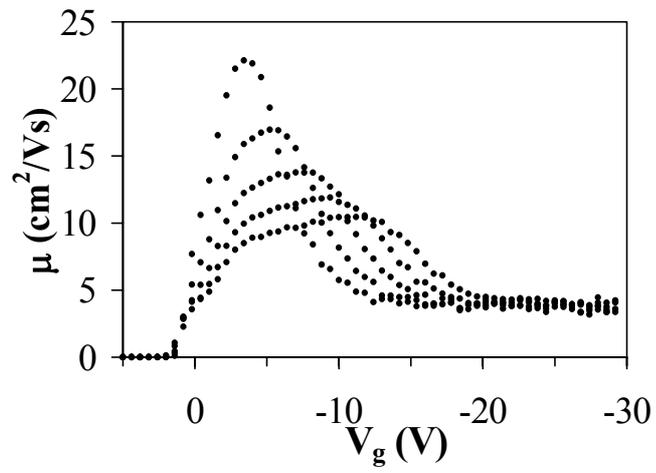

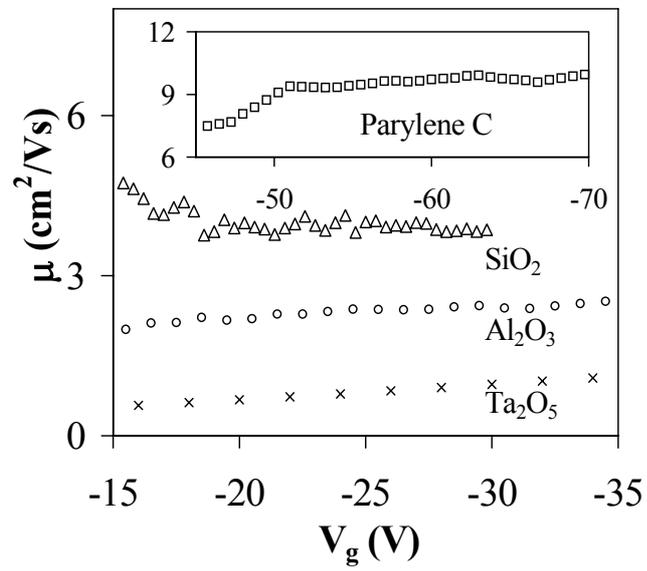

Figure 3 A. F. Stassen *et al.*

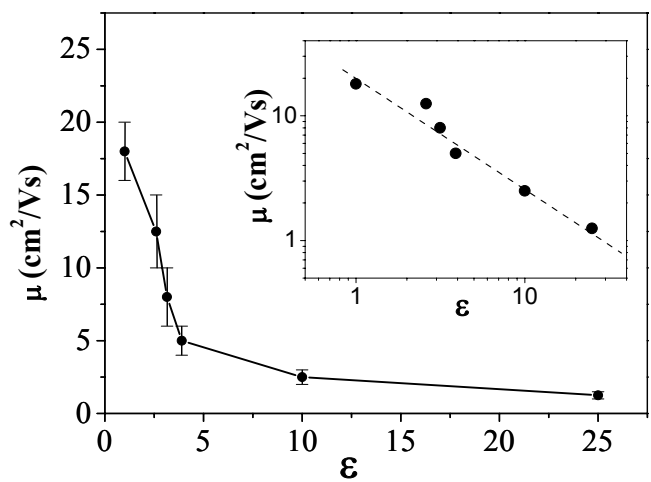

Figure 4 A. F. Stassen *et al.*